\documentclass[aps,pre,superscriptaddress,showkeys,reprint]{revtex4-2}
\usepackage{amsmath,amssymb}
\usepackage{graphicx}
\usepackage{dcolumn}
\usepackage{bm}
\usepackage[latin1]{inputenc}
\usepackage{float}
\usepackage{color}
\usepackage{booktabs}
\usepackage{amsmath}
\usepackage[toc,page]{appendix} 
\usepackage{latexsym}
\usepackage{amsmath}
\usepackage{mathrsfs}
\usepackage{bm}
\usepackage{amsthm}
\usepackage{physics}
\usepackage{adjustbox}
\usepackage{bbm}
\usepackage{amsfonts}
\usepackage{amssymb}
\usepackage{color}
\usepackage{epsfig}
\usepackage{hyperref}
\hypersetup{colorlinks=true, citecolor=blue}
\usepackage{soul}
\usepackage{cancel}
\usepackage[normalem]{ulem}
\usepackage[cmtip,all]{xy} 
\usepackage{comment}
\usepackage{multirow}
\usepackage{mathtools}
\usepackage{dsfont}

\begin{document}

\title{Minimal Model for Chirally Induced Spin Selectivity: Chirality, Spin-orbit coupling, Decoherence and Tunneling}

\author{Miguel Mena}
\affiliation{Departamento de F\'isica, Colegio de Ciencias e Ingenier\'ia, Universidad San Francisco de Quito, Diego de Robles y V\'ia Interoce\'anica, Quito, 170901, Ecuador}

\author{Solmar Varela}
\affiliation{Escuela de F\'isica, Universidad Central de Venezuela, UCV. C\'odigo Postal 1050. Caracas, Venezuela}

\author{Bertrand Berche}
\affiliation{Laboratoire de Physique et Chimie The\'oriques, Universit\'e de Lorraine, CNRS, Nancy, France}

\author{Ernesto Medina}
\affiliation{Departamento de F\'isica, Colegio de Ciencias e Ingenier\'ia, Universidad San Francisco de Quito, Diego de Robles y V\'ia Interoce\'anica, Quito, 170901, Ecuador}

\date{\today}

\begin{abstract}
Here we review a universal model for chirally induced spin-selectivity (CISS) as a standalone effect occurring in chiral molecules. We tie together the results of forward scattering in the gas phase to the results for photoelectrons in chiral self-assembled monolayers and the more contemporary results in two terminal transport setups. We discuss the ingredients that are necessarily present in all experiments to date, which we identify as: i) chirality, be it point, helical or configurational, ii) the spin-orbit coupling as the spin active coupling of atomic origin, iii) decoherence as a time-reversal symmetry breaking mechanism that avoids reciprocity relations in the linear regime and finally iv) tunneling that accounts for the magnitude of the spin polarization effect. This proposal does not discard other mechanisms that can yield comparable spin effects related to interactions of the molecule to contacts or substrates that have been proposed but that are less universal or apply to particular situations. Finally, we discuss recent results suggesting CISS as a molecular phenomenon in the real of enantiomer selectivity, coherent electron transfer, and spin effects in chiroptical activity. 

\end{abstract}


\maketitle

\section{Introduction}
Spin activity in low-dimensional systems, such as molecules, has become a topic of high interest since the work of Farago\cite{Farago1980,Farago1981}. Farago found that oriented molecules with point chirality (chiral centers) interacted with electron spin through the spin-orbit (SO) coupling, producing an angular-dependent spin polarization. Hegstrom\cite{ Hegstrom1982} pointed out that chiral effects are also expected in dissociation and rearrangement collisions. Following Kessler's group found these effects in randomly oriented point chiral molecules in the gas phase\cite{Kessler1995,Kessler1996} doped with heavy atoms to enhance the SO strength. The polarizing effect, although measurable, was very small, i.e., $~10^{-4}\%$. 
\begin{figure}[h!]
    \centering
    \includegraphics[width=8cm]{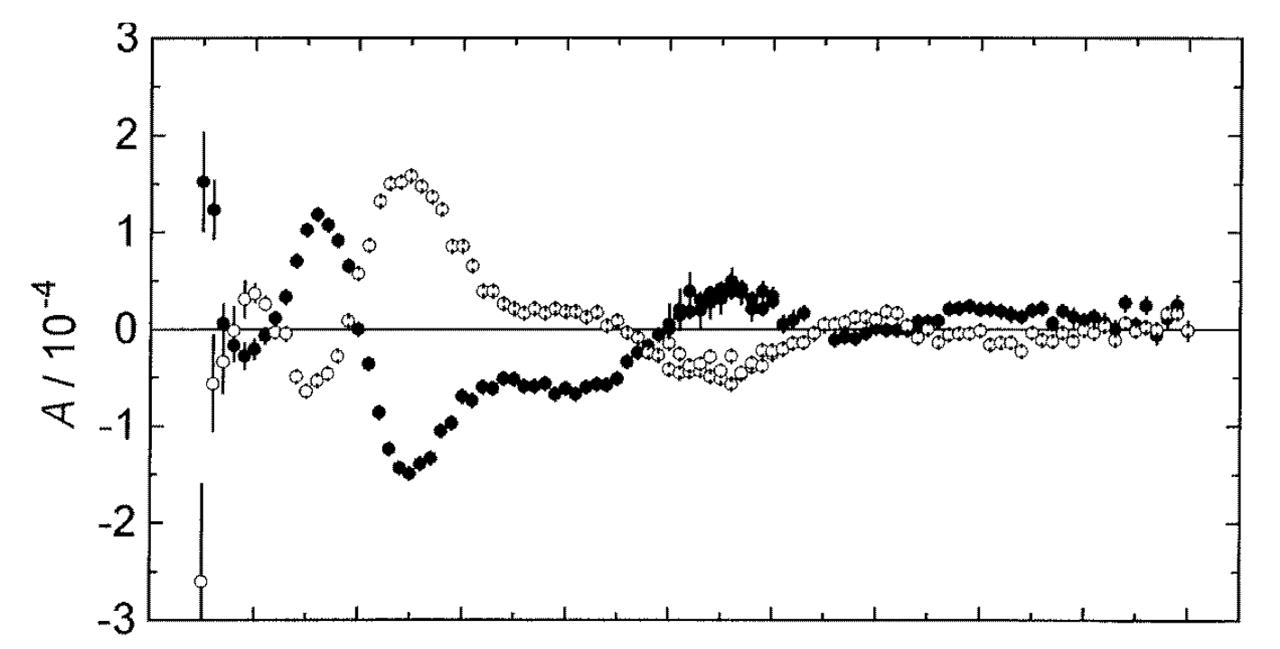}
    \center
    \includegraphics[width=8cm]{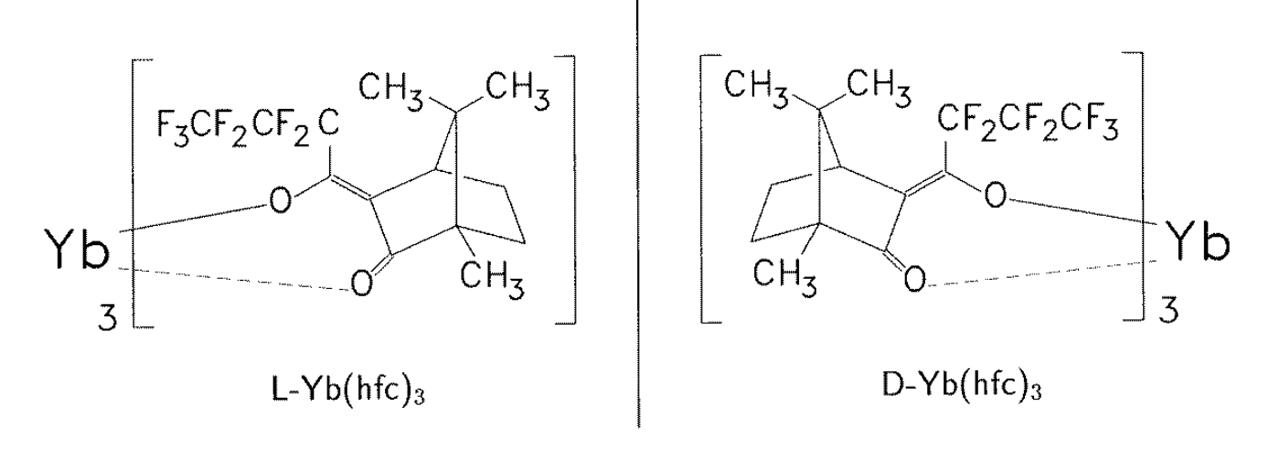}
    \caption{Electron spin transmission asymmetry  through gaseous phase of ${\rm D-Yb(hfc)}_3$ (black circles) and and L-Yb(hfc)3 (white circles) chiral molecules. The bottom panel shows the target chiral molecules featuring a heavy atom for strong SO coupling producing by itself no spin selective forward scattering. From reference \cite{Kessler1996}.
    }
    \label{KesslerThompsonBlum}
\end{figure}

Later developments\cite{BlumThompson1997} showed that arrays of oriented chiral molecules could, in fact, increase the electron polarizing effects by two to three orders of magnitude, putting the polarization efficiency at ~0.01\%. A very detailed theory of the scattering process that encompassed chiral and oriented molecules was articulated by Johnston, Blum, and Thompson\cite{BlumThompson1993,BlumThompson1997}. In reference \cite{BlumThompson1989} Blum and Thompson derived explicit physical requirements for polarization of forward scattering electrons for both planar-oriented and point chiral molecules. They found that spin flipping amplitudes are equal $|f(1/2,-1/2)|^2=|f(-1/2,1/2)|^2$ and spin polarization only depends on the interference between forward unscattered electrons and the amplitudes $f(1/2,1/2)$ and $f(-1/2,-1/2)$. Only chiral molecules can polarize in non-oriented mixtures, while the non-chiral mixtures average out spin polarization to zero. It must be clarified that the reported polarization is only for forward scattering, and there is actually a spin-polarized angular distribution whose averaged spin polarization is null since time-reversal symmetry is in place. For the second Born approximation/double scattering, the results for the angular dependence of the spin polarization and inelastic scattering can be found in references \cite{MedinaLopez2012,VarelaLopez2014,Paltiel2020}.

New developments followed soon after, with self-assembled mon/multilayers (SAM) of amino acids (point chirality) on metallic surfaces\cite{NaamanScience}. Unpolarized photoelectrons emerging from the surface were polarized between 10-20\%, three to four orders of magnitude stronger than gas-phase chiral molecules. The same setup with helical structures, such as double-stranded DNA\cite{Xie2011}, yielded even higher longitudinal polarizations of up to 60\% as measured of the emerging free electrons by a Mott detector. Such high polarizations are unheard of even for electrons emerging from ferromagnets, the basis technology for giant magnetoresistance spintronics.  These measurements can, in principle, be well described by the forward scattering formalism of Blum and Thompson for the case of oriented chiral molecules from the symmetry point of view. In particular, the target molecule is more complex and should be treated in a higher-order Born series. The approach of Blum and Thompson already bears atomic spin-orbit interactions.

Doubts about whether the chiral SAMs were a collective phenomenon or a single molecule effect suggested materializing single molecule circuits, e.g., molecules attached on one end to a metallic or ferromagnetic surface while on the other to a metallic nanoparticle attached to a contact AFM measurement device\cite{Xie2011}. This setup proved that, in fact, this effect was a single molecule effect with no apparent collective mechanisms. By then, the effect was named chirally induced spin selectivity or CISS\cite{ACSNanoReview}. A large body of convincing experiments followed, consolidating the spin-polarizing phenomenon that includes polarization by oligopeptides, alpha-helices in proteins, and photosystem I, among others, testing different measurement principles, elucidating possible effects of the contacts and the linker molecules used to connect to different surfaces.

The theoretical approaches to CISS were recently reviewed in ref.\cite{Matiyahu2016}. Models have been developed from tight-binding approaches inspired directly by partially filled orbital structures of chiral molecules\cite{Yeganeh2009,VarelaMujicaMedina2016,Oligopeptides2020,PeraltaFeijoo2020,PeraltaFeijoo2023} or by the discretization of continuum models containing the canonical SO term derived from the non-relativistic limit of the Dirac equation\cite{GuoSun2014,GutierrezDiaz2012,Matiyahu2016}. The approaches based on orbital coupling have the advantage of identifying the magnitude of the transport couplings and their origin. This is how SO magnitudes were established to be in the meV range. The second approach relies on fitting, e.g., spin active ingredients, to the transport measurements. An even more efficient approach was developed by Kochan et al. in \cite{Kochan}, where group theory according to the spatial group, combined with the identification of the corresponding matrix elements. We then have the rigor of completeness of possible spin active interactions and their magnitudes. The previous exercise assessed the spin active ingredients in the modeling and their relative presence in chiral molecules. This has not been addressed extensively in the literature.

The two terminal setups have been the most explored experimentally for CISS. A realization that stirred the theoretical community around this topic is that most models were time-reversal invariant so that reciprocity dictates that two terminal setups cannot polarize spin\cite{YangVanWees2020}. The SO coupling breaks space inversion symmetry but preserves time-reversal symmetry. These symmetries yield helicity states where the eigenstates have definite spin projections onto the momentum directions\cite{MedinaGonzalez,Guinea}. Reciprocity requires that $G(M)=G(-M)$ so the detector on one end endowed with a magnetization $M$ could not distinguish any polarization produced by the chiral molecule\cite{YangVanWees2020}. The general rule does not distinguish whether the two terminal devices have one or more channels. A version of the reciprocity rules for a single channel of the two-terminal setup is Bardarson's theorem, or the so-called ``single channel no go theorem," which applies to non-interacting wires in the linear regime\cite{Bardarson2008}. 

\begin{figure}[h]
    \centering
    \includegraphics[width=7cm]{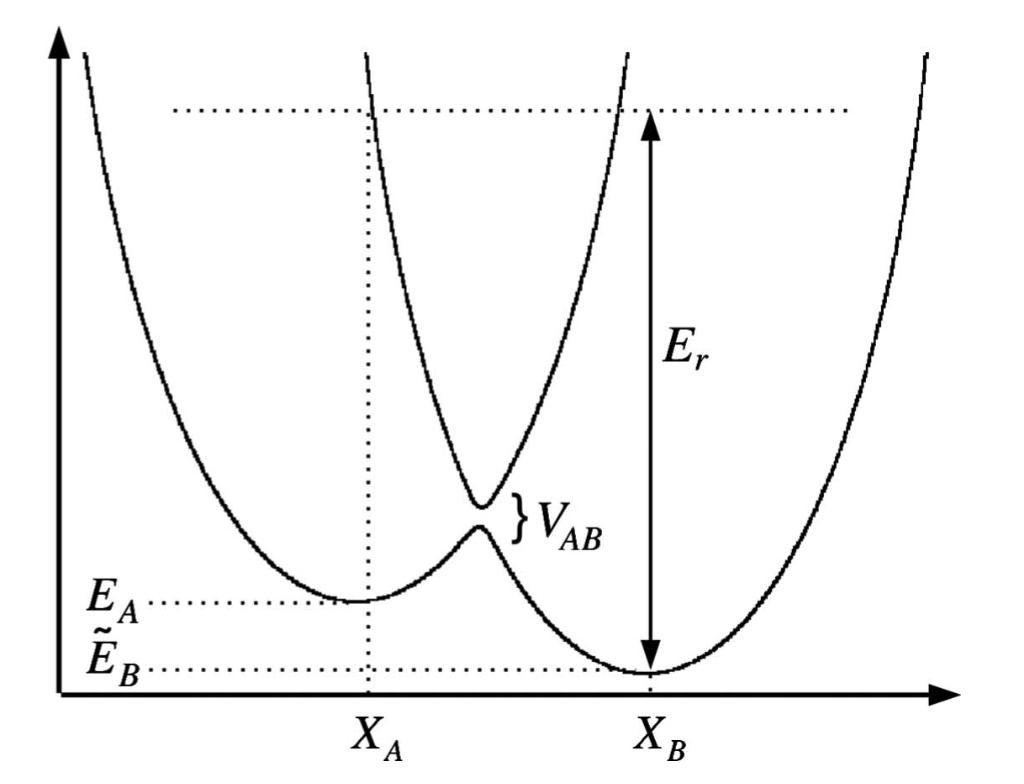}
    \caption{Marcus-Hush model for electron transfer reactions between $X_A$ and $X_B$ with a tunneling barrier $V_{AB}$. $E_r$ is the reorganization energy due to the electron-phonon coupling. From ref. \cite{PastawskiCattena2010}.}
    \label{FigCattenaPastawski}.
\end{figure}

A particularly physically appealing way to circumvent reciprocity in the linear regime that applies to almost all experiments is to effectively add a third probe to the molecule. This is conceptually identical to the standard linear electron-phonon coupling used to describe the Franck-Condon effect and the electron-transfer process\cite{PastawskiCattena2010,Book1,Book2} depicted in Fig.\ref{FigCattenaPastawski} where the coupling to the reservoir via a linear electron-phonon coupling that determines a reorganization energy $E_r$. The model assumes a tunneling energy $V_{AB}$ to transfer electrons from $A\rightarrow B$. The resulting rate is then\cite{PastawskiCattena2010}
\begin{equation}
k_{A\rightarrow B}=\frac{1}{\tau_{A\rightarrow B}}=\frac{2\pi}{\hbar}|V_{AB}|^2F(\Delta E),
\end{equation}
where $F(\Delta E)$ (with $\Delta E=E_A-\tilde E_B$), the Frank-Condon factor, has a Gaussian energy dependence. One thus includes an essential ingredient to transport and electron transfer in molecules, particularly chiral molecules.

The first model to effectively include a third probe that broke time-reversal symmetry in the context of CISS was Guo and Sun\cite{GuoSunPRL}. In fact, without a decoherence probe, they could not see any polarization effect, no matter how strong the SO coupling. Also, their model revealed the importance of chirality, either explicitly as in chiral centers or helical structures or induced by terminal/scattering configuration\cite{GuoSunNanotube,Paltiel2020}. The scattering configuration ``chirality" was pointed out in reference \cite{BlumThompson1989} through a solid symmetry analysis of non-intrinsically chiral structures depending on their orientation with respect to the electron momentum direction. In particular, the case of planar molecules like water with a definite orientation to the electron $\bf k$ vector produces a higher forward scattering amplitude for the up spin than for the down spin, having chosen the propagation axis as the quantization direction. An argument comparing the screw direction of spin and the screw direction of the target clearly distinguishes spin scattering preference that applies to the achiral nanotube case studied in reference \cite{GuoSunNanotube}. Later work \cite{Matiyahu2016,Huisman,Fransson} also recognized that one could bypass the reciprocity requirements of two terminal devices by introducing decoherence through an imaginary component to site energies at third probe sites.

\begin{figure}[h]
    \centering
\includegraphics[width=7cm]{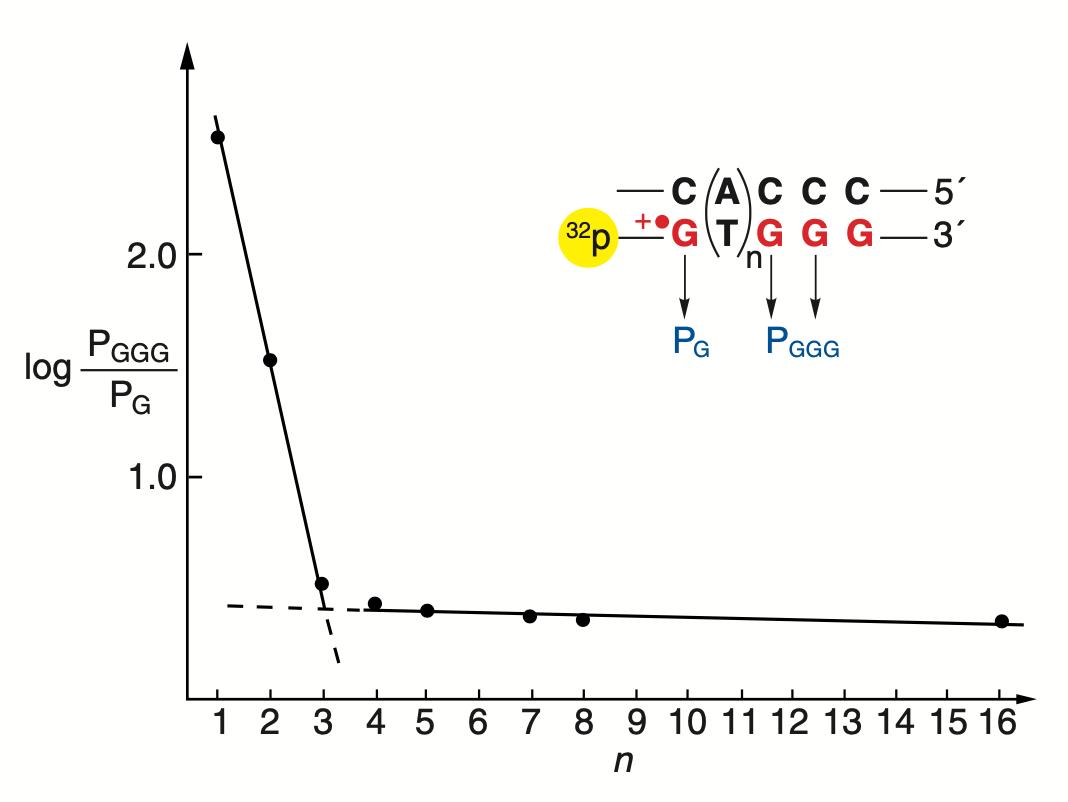}
    \caption{Fitting of electron transfer measurements between a GGG sequence and a guanine cation separated by a sequence of $n$ AT pairs. The efficiency of the transfer is expressed in terms of the ratio of the relative presence of electrons on either side of the AT sequence. From reference \cite{Giese2002}.}
    \label{FigGiese}
\end{figure}

The previous model, adding decoherence to SO coupling and chirality, in fact, opens many connections to the successful description of electron transfer mechanisms in molecules, starting from the Frank-Condon effect mentioned before and more general polaron transport mechanism \cite{PastawskiCattena2010,BarrosoDiaz2022}. Maybe the most emblematic electron transfer dependence on length for DNA is that obtained by Giese \cite{Giese2002} where there is a transition between a clear tunneling dependence to a weak power-law decay (see Fig.\ref{FigGiese}).

In fact, many molecular wire-type molecules follow this law, as pointed out by ref.\cite{Lindsay2020} for molecular wires such as oligophenylineimide, peptides, and even complex structures such as antibodies. This length dependence loosely indicates a transition from tunneling to a hopping behavior that has often attempted to be modeled in various ways; see, for example, ref.\cite{GutierrezCuniberti}.
\begin{figure}[h]
    \centering
    \includegraphics[width=6.5cm]{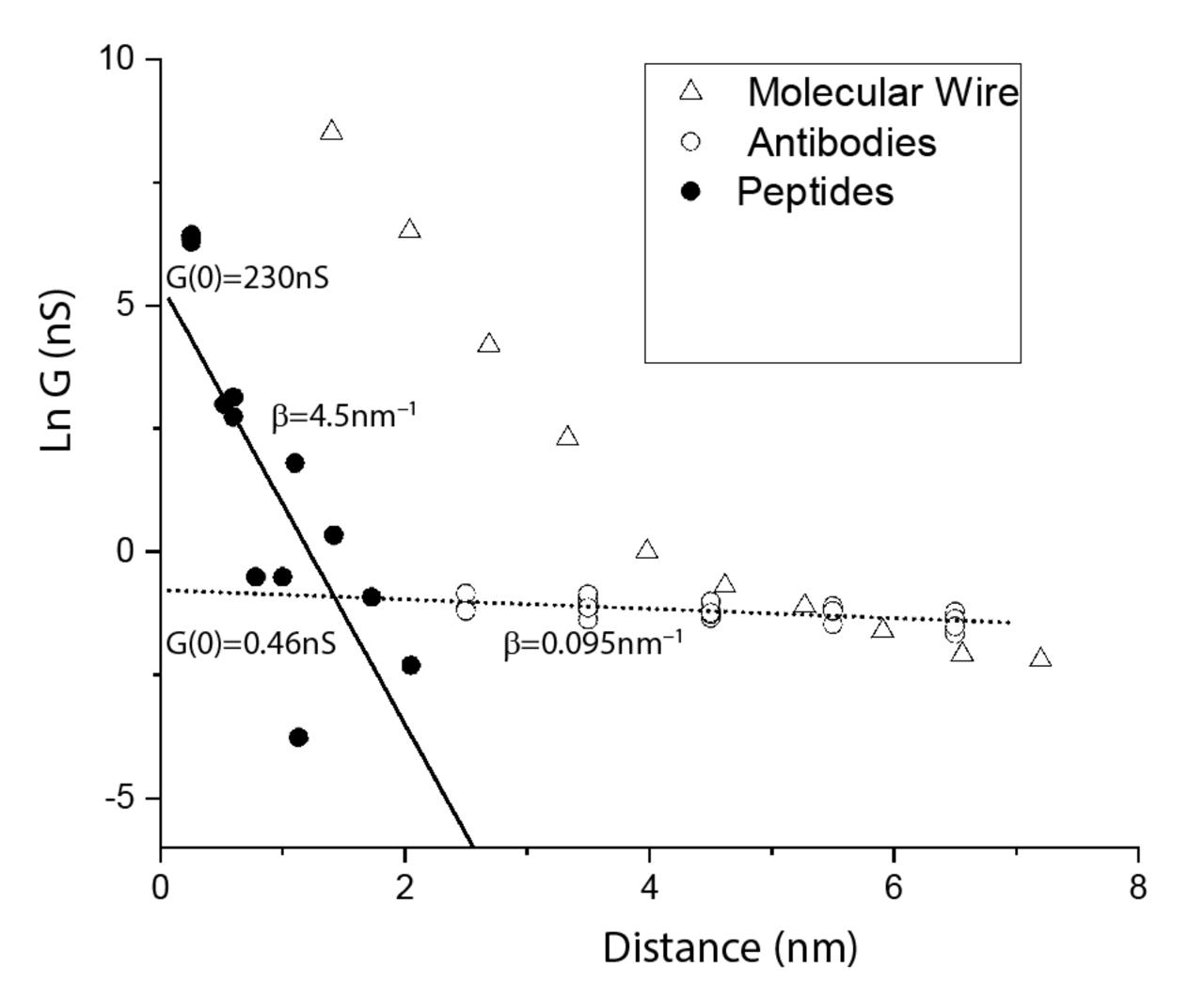}
    \caption{Behavior of conductance for complex molecular wires such as oligophenylineimide, peptides and antibodies. The $\beta$ reported corresponds to an exponential fitting. From reference \cite{Lindsay2020}.
    }
    \label{FigLindsay}
\end{figure}

\begin{figure}[h]
    \centering
    \includegraphics[width=4.5cm]{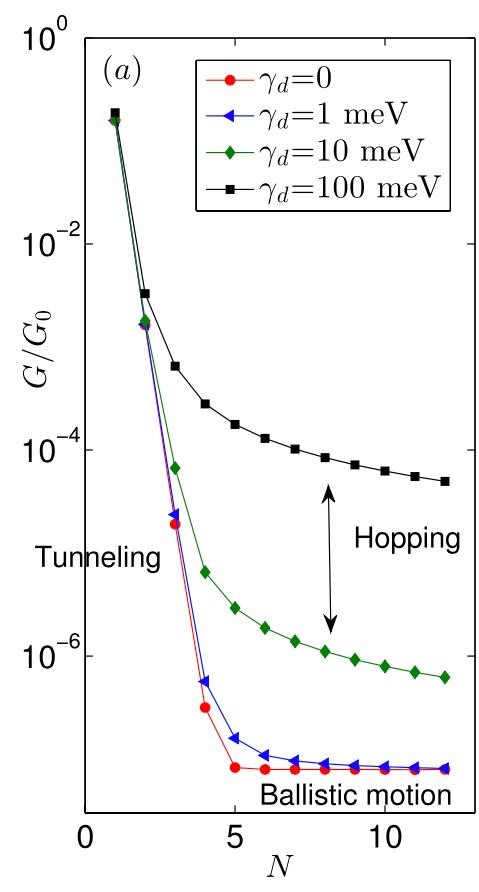}
    \caption{Tunneling to hopping transition through a decoherence model probe as a function of length. Different curves depend on the coupling $\gamma_d$ of the D'Amato-Pastawski probe to the DNA model. From reference \cite{Kilgour}.}
    \label{ModelingDamatoPastawski}
\end{figure}

A convincing model of this behavior was presented recently by Kilgour and Segal\cite{Kilgour} using the D'Amato Pastawski Hamiltonian probe. Figure \ref{ModelingDamatoPastawski} shows how, via the introduction of decoherence events coupled by an energy $\varepsilon$ dependent probe $\gamma_d=2{\tilde\gamma}_d\Omega|\varepsilon|/(\varepsilon^2+\Omega^2)=\Im\Sigma_d$, it is possible to explore transitions from the tunneling regime to the hopping or ballistic motion.

\section{Decoherence as a time reversal breaking mechanism}

\begin{figure}[h!]
    \centering
    \includegraphics[width=8cm]{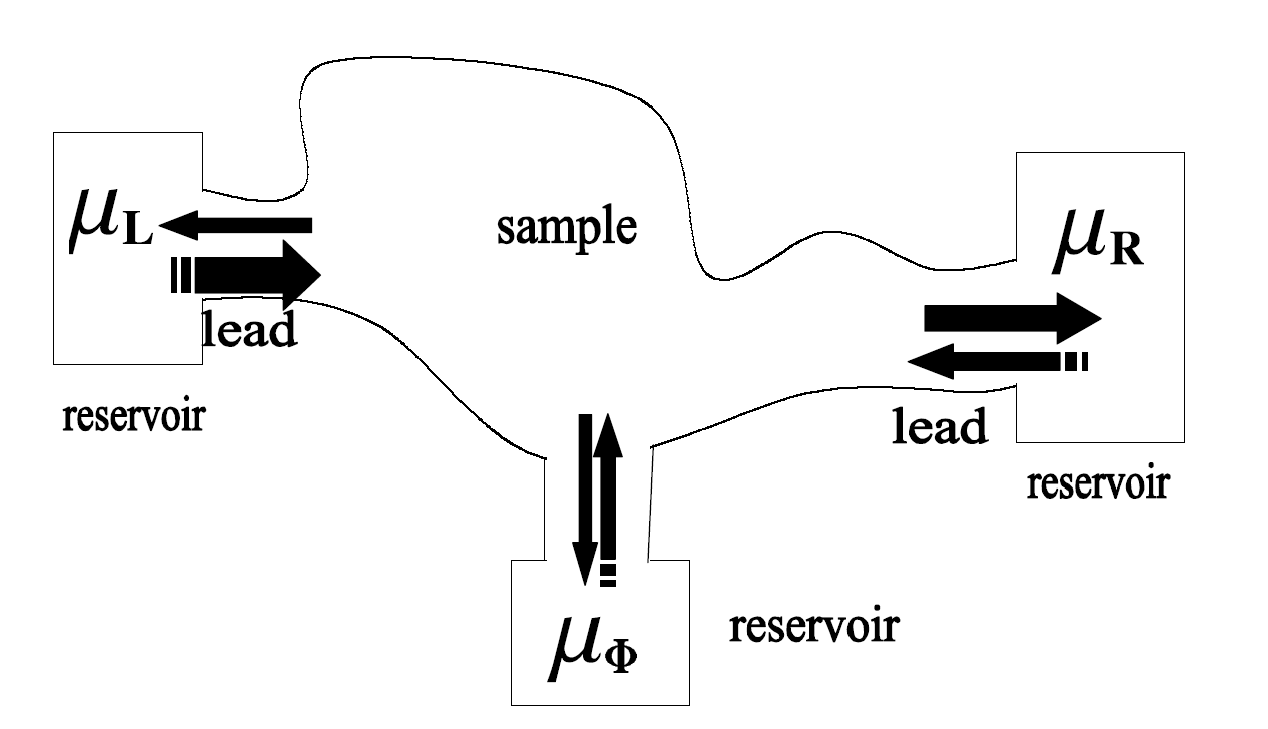}
    \caption{Two terminal setup with voltage probe $\mu_{\phi}$. The condition of $I_{\phi}=0$ preserves unitarity while decoherence is induced by loosing phase memory between input and output amplitudes. From ref.\cite{RevMex}.}
    \label{RevMexicana17}
\end{figure}

If time reversal symmetry is broken, then the prohibition of spin filtering in a two-terminal setup in the linear regime is mute.  Decoherence is a general way to avoid reciprocity restrictions (nevertheless see ref.\cite{Utsumi}) and should be involved in all two-terminal experimental setups to date. How to incorporate the action effects of reservoirs coupled to a low dimensional system in the transport configuration has a long history and is well understood\cite{ButtikerProbe,DAmatoPastawski}. For a more generic context, see ref.\cite{Schlosshauer}. We will briefly describe the Hamiltonian approach to decoherence in transport of D'Amato and Pastawski\cite{DAmatoPastawski}. Buttiker's approach is equivalent but expressed in terms of scattering matrices. A basic understanding of the two terminal setups, i.e., two contact reservoirs with an additional decoherence probe/coupling to the environment (see Fig.\ref{RevMexicana17}), is given by the following system of equations
\begin{widetext}
\begin{equation}
\begin{pmatrix}
I_L \\ 
I_{\phi} \\ 
I_R 
\end{pmatrix}= 
\begin{pmatrix}
    -(G_{R,L}+G_{\phi,L}) & G_{L,\phi} & G_{L,R}\\
    G_{\phi,L} & -(G_{R,\phi}+G_{L,\phi}) & G_{\phi,R}\\
    G_{R,L} & G_{R,\phi} & -(G_{\phi,R}+G_{L,R})
\end{pmatrix}
    \begin{pmatrix}
        \mu_L\\
        \mu_{\phi}\\
        \mu_R
    \end{pmatrix}
\end{equation}
\end{widetext}
where $I_{L,\phi,R}$ are the currents at the left, decoherence lead, and right, $\mu_{L,\phi,R}$ the corresponding potentials, and  $G_{L,R}$ is the Green's function connecting the left and right terminals. In contrast, the $G_{L,\phi}$ connects the left terminal and the decoherence site/voltage probe site. The general structure gives the Green's function between two sites $I,J$
\begin{eqnarray}
    G_{I,J}^{R}&=&\langle I|[\varepsilon \hat I-\hat H]^{-1}|J\rangle\nonumber\\
    &=&\frac{V_{I,J}}{[\varepsilon-({\tilde E}_I+\Sigma_I)][\varepsilon-({\tilde E}_J+\Sigma_J)]-V_{I,J}V_{J,I}}.
\end{eqnarray}
Here $H_{\rm eff}$ is the effective Hamiltonian of the system\cite{RevMex,FoaPastawski}, $V_{I,J}$ is the normalized coupling between those sites, and ${\tilde E}_I$ is the normalized energy at site $I$. Finally, $\Sigma_I$ is the self-energy of a decoherence center/voltage probe at site $I$.

The site for decoherence $\phi$ is generally characterized by a site energy in the full Hamiltonian plus a complex self-energy associated with an external reservoir. If one sets the $I_\phi=0$, we have a voltage probe and decoherence effects that preserve unitarity. On the other hand, if $I_{\phi}\ne 0$, we have probability leakage and thus a nonunitary source of decoherence\cite{GuoSun2014,Matiyahu2016}. Buttiker's probe is a voltage probe preserving unitarity, so it belongs to the former case.

In the unitary case, if we solve the current equation as a function of the terminal voltages, we arrive at the equation for the equivalent conductance ${\cal G_{I,J}}=(e/h) T_{I,J}$
\begin{equation}
\tilde{\cal G}_{R,L}={\cal G}_{R,L}+({\cal G}_{R,\phi}^{-1}+{\cal G}_{\phi,L}^{-1})^{-1},
\end{equation}
where the second term is an ohmic term due to the decoherent effect of the lateral reservoir. One can see that the decoherent site opens up a second channel for transport, even in one-dimensional systems, avoiding the premises of Bardarson's theorem\cite{Bardarson2008}. The irreversible character of the second term breaking phase coherence in the system breaks time-reversal symmetry, and the system is no longer subject to two terminal reciprocity. Generalizations to many voltage probes can be implemented systematically. For details, see ref.\cite{RevMex,FoaPastawski}; also, perturbative treatment of decoherence can be formulated.

The second form of decoherence is leakage or loss of unitarity in the transmission, which is much easier to implement in the model. As a reference model, a semi-infinite lead is used, whose self-energy can be exactly calculated analytically in terms of the lead tight-binding parameters. This gives a single-band picture of the reservoirs with a semicircular law for the density of states.

The latter lends a general picture of how to break time-reversal symmetry in the transmission problem. However, how this happens physically, in the one-body picture, is through the coupling of transport electrons to excitations through the electron-phonon or spin-phonon coupling, or the electron-electron interaction. One then must also include the thermal populations of these coupled excitations to compute specific signatures to transport through temperature or energy dependences.

In the context of molecular vibrations, recent work of Cattena and Pastawski\cite{PastawskiCattena2010} has derived from microscopic torsion modes the relaxation times of transport electrons in terms of the imaginary part of the self energies involving the coupling to the thermalized environment of torsional modes. Through the Fermi golden rule, they connect the relaxation time to torsional vibrations self energies as (see also \cite{RevMex})
\begin{equation}
    \tau_{\rm ph}=\frac{\hbar}{2}\frac{1}{\Im \Sigma_{\rm el-ph}}.
\end{equation}
A recent article analyzes model electron-phonon, electron impurity, and electron-electron interaction, which have been discussed in the context of theoretical models of CISS, arriving at the self energies\cite{Kemper2018}
\begin{eqnarray*}
    \Im \Sigma_{\rm el-ph}^R(t,t)&=&-g^2[2 n_B(\Omega/T)+1],\nonumber\\
    \Im \Sigma_{\rm el-imp}^R(t,t)&=&-V^2,\nonumber\\
    \Im \Sigma_{el-el}^R)(t,t)&=&-{\bar U}^2n(1-n)\equiv -U^2,
\end{eqnarray*}
where $\Omega$ is the characteristic Einstein phonon frequency coupled to electrons, and $n_B$ is the phonon occupation. The electron density per spin defines $n=\langle n_{\downarrow}\rangle=\langle n_{\uparrow}\rangle$, $U$ is the Hubbard coupling, and $V$ is the coupling to impurity levels.  Each type of self-energy should imprint on particular features of CISS spin filtering as a self-energy signature. Then, a complete, well-established formalism incorporates decoherence into transport via interactions.

We have now surveyed three minimal ingredients for CISS to operate as an intrinsic effect in the two terminal transport setups in the linear regime: Chirality, spin-orbit coupling, and decoherence. Chirality is particularly coupled to the SO coupling since it provides the inversion asymmetry that makes it non-zero\cite{VarelaMujicaMedina2016}. It is also clear that the main source of SO coupling is the atomic coupling\cite{VarelaMujicaMedina2016}.  Decoherence, however, makes for the time-reversal symmetry breaking that enables two-terminal spin polarization. As refs.\cite{GuoSunPRL,YangVanWees2020} first demonstrated, no time-reversal symmetry breaking means reciprocity disallows spin polarization. 

\section{Decoherence and spin polarization}
The previous evidence is compelling from a symmetry point of view, but it remains unclear how specifically spin-blind decoherence produces spin polarization. The greatest misgivings of the theory have not been to find ways in which spin polarization can occur but to predict the large spin polarization of more than ~50\% that is observed experimentally (see, e.g., ref.\cite{Xie2011,NickelPolarizationNaaman}). The first models needed to adjust the SO magnitude to produce a few percent effects\cite{GutierrezCuniberti,GuoSunPRL,Herrmann,Matiyahu2016}. Generally, these couplings were not justified by the electric fields available in the molecular systems involved. Some efforts were directed to finding out whether molecular vibrations could be exposing the electric fields of atomic nuclei, weakly violating the Born-Oppenheimer and thus enhancing the atomic SO coupling\cite{Subotnik} beyond the meV range \cite{Simserides,MedinaLopez2012,VarelaMujicaMedina2016}. 

Seeking an additional ingredient that can explain the size of the effects and is also generic to most of the experiments exhibiting CISS brings us to tunneling electron transfer. Tunneling was included as a necessary part of physics by Michaeli and Naaman \cite{Michaeli}, where time-reversal symmetry breaking was introduced by a magnetic field presumably induced by the curvature of the helical structure. Nevertheless, such curvature is not a generic element of chiral systems, such as point chiral molecules where CISS has been observed.

Here, we review the simplest transmission model through an SO active barrier and discuss the interplay between the minimal ingredients advocated for CISS. We explore the consequences of the exactly solved model in ref.\cite{ScipostTunneling} for spin-polarized electron transmission in a one-dimensional two-probe setting. The action of a $U(1)$ field on tunneling electrons\cite{Buttiker} under a barrier relaxes the electron spin toward the lower energy orientation. This behavior is contrasted with the effective momentum-dependent magnetic field arising from the SO coupling.

\subsection{Tunneling and spin relaxation in a magnetic field}

We have covered the elements of chirality, spin-orbit coupling, and, finally, decoherence as reciprocity-breaking. From the point of view of symmetry, these ingredients do not allow us to estimate the degree of resulting spin polarization. In fact, previous models that achieved the experimental degree of polarization assumed oversized spin active parameters.  This is due to models that could not be connected to the relevant orbital mechanisms producing the spin-orbit coupling. The models that included both orbital mechanisms and their filling were considered first by ref.\cite{VarelaMujicaMedina2016,Oligopeptides2020} for DNA and Oligopeptides, and these models conclude that, as in semiconductors\cite{Winkler}, the SO is of atomic origin, and in the meV range from carbon nitrogen atoms. With this value, the previous models all produce polarizations ten to a hundred times smaller than the experiments. To bridge this gap, there is a universal fact in electron molecular transport/transfer, i.e. {\it tunneling}, be it between terminals (electron transport) or internal sites of the molecule (electron transfer)\cite{Lindsay2020}. 

A very illuminating model for the effects of a barrier on spin polarization is due to Buttiker\cite{Buttiker} and illustrates the importance of breaking time reversal to obtain spin polarization by relaxing to the direction of the magnetic field.  The solution to this problem allowed for properly addressing the tunneling time problem by realizing that the spin-dependent decay of the wavefunctions under the barrier modulates spin precession in the field. We will summarize here the findings in ref.\cite{ScipostTunneling}.  The Hamiltonian, in this case, is given by
\begin{equation}
    \cal{H}=\begin{cases}
			(\frac{p_x^2}{2m}+V_0)\mathds{1}_{\sigma}-\Gamma \sigma_z, & \text{if~ $0<x<a$}\\
            (\frac{p_x^2}{2m})\mathds{1}_{\sigma}, & \text{otherwise},
		 \end{cases}
\label{HamiltonianMagneticField}
\end{equation}
where $\mathds{1}_{\sigma}$ is the unit matrix in spin space and $\sigma_i$ are the Pauli spin matrices, with $i=x,y,z$. $\Gamma=\hbar\omega_L/2$ where $\omega_L$ is the Larmor frequency, {\color{black}and $V_0$ is the barrier height}. $\cal{H}$ acts on the spinors $\psi=\left(\psi_+(x)~\psi_-(x)\right)$. The magnetic field only acts under the barrier, breaking time-reversal symmetry in the tunneling process. 

The {incoming} wavefunction was chosen to be
\begin{equation}
    \psi=\frac{1}{\sqrt{1+|s|^2}} \begin{pmatrix}1\\{i s}\end{pmatrix}.
    \label{waveBField}
\end{equation}
The values of $s=\pm 1$ correspond to the two eigenfunctions of the $\sigma_y$ matrix, and $s=\pm i$ correspond to the two eigenfunctions of the $\sigma_x$ matrix, appropriately normalized. This form is sufficiently general for the purposes of showing spin relaxation. The eigenvalues of the Hamiltonian are $E=\frac{p_x^2}{2m}-\sigma\Gamma +V_0$,
where $\sigma=\pm 1$ is the spin degree of freedom. Using $E=\hbar^2 k^2/2m$, we define the wavevector outside the barrier $k$. From the eigenvalue equation, one can then distinguish between the different wavevectors under the barrier $\kappa_{\sigma}^{\lambda} =\lambda\left(k^2-k_0^2+\sigma k_B^2\right)^{1/2}$
where $k_B^2=2m\Gamma/\hbar^2$ and $k_0^2=2m V_0/\hbar^2$. As can be seen, $\kappa_{\sigma}^{\lambda}$ can only be real or imaginary. Thus, we have either exponentially decaying solutions for $k^2<k_0^2-\sigma k_B^2$ or plane waves otherwise.
The boundary value problem was solved by taking care of flux conserving conditions\cite{ScipostTunneling}.

When there is no barrier present, only Larmor precession ensues, and no distinction occurs between spin species. On the other for energies under the barrier, the transmission with the length of the barrier obeys $T_+\sim e^{-2\kappa_+^+ a}$, and $T_-\sim e^{-2\kappa_-^+ a}$ as long as $E<V_0\mp\hbar\omega_L/2$ marking a different decay for each spin species and thus a polarizing effect of the field. The expectation value for the $i$-th polarization component of the electron for the transmitted (T) wave is given by
$\langle s_i\rangle=\frac{\hbar}{2}\langle\psi_T|\sigma_i|\psi_T\rangle$ 
for the transmitted wave. 
\begin{figure}
    \centering
    \includegraphics[width=8cm]{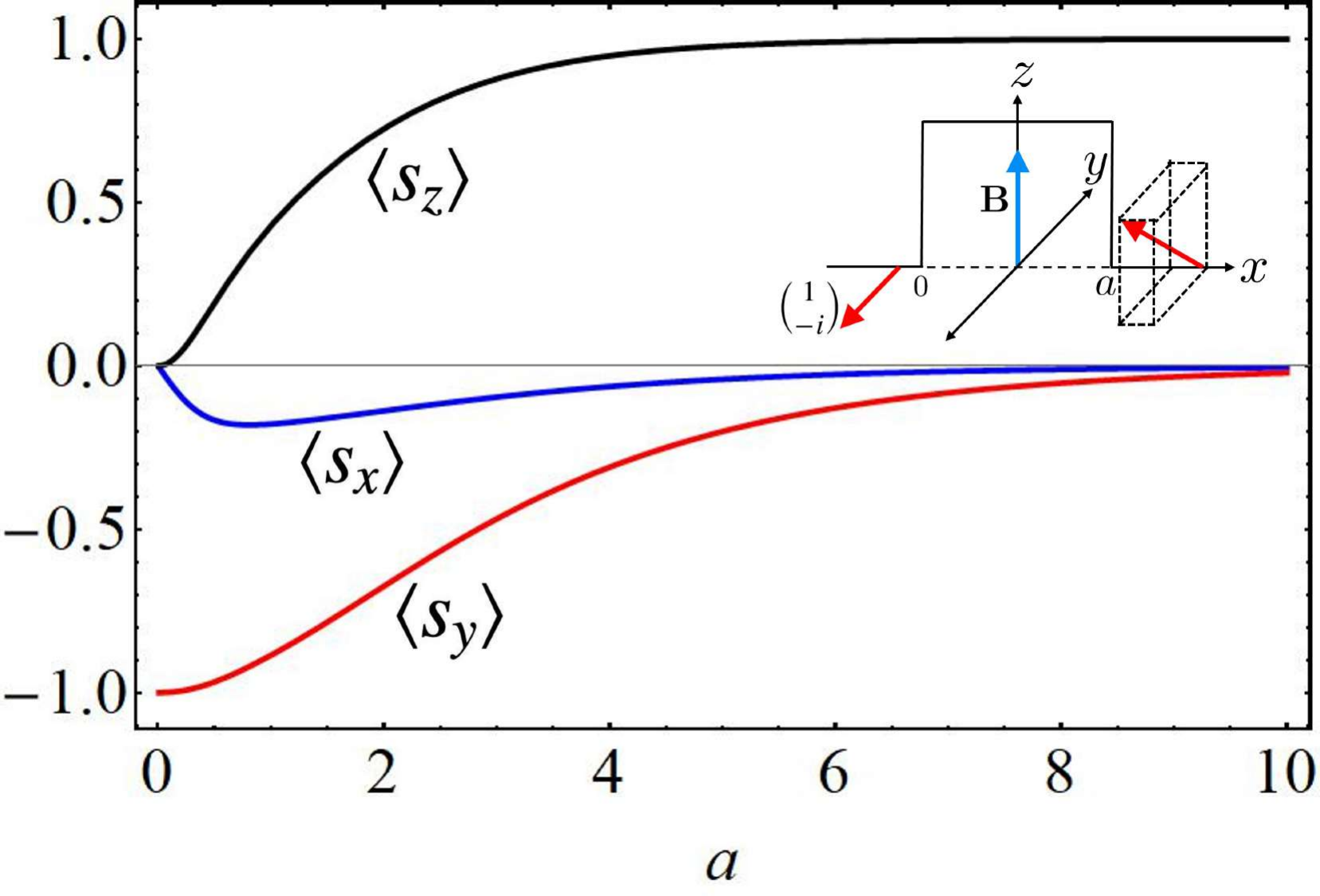}
    \caption{Spin relaxation toward magnetic field direction due to tunneling when $k^2<k_0^2\mp k_B^2$\cite{Buttiker}. Polarization increases exponentially with the tunneling length, making for a very efficient spin polarizer. The reference values taken for the plot are $k=2/a$, $k_0=3/a$, and $k_B=1/a$, where $a$ is the barrier length. From ref.\cite{ScipostTunneling}.}.
    \label{LarmorUnderBarrier}
\end{figure}
On the other hand, for $V_0 > E\pm\hbar\omega_L/2$, spin precession around the magnetic field is only part of the average spin motion since each spin component decays at a different rate under the barrier. This gives rise to a $z$-component that aligns with the direction of the field\cite{Buttiker}. Figure \ref{LarmorUnderBarrier} from ref.\cite{ScipostTunneling} depicts the qualitative motion for the latter case. The combination of time symmetry breaking through an external magnetic field and a tunneling barrier makes for a very strong spin polarizer.

\subsection{Spin-orbit coupling plus decoherence as an effective magnetic field}
As a reference model for spin-orbit coupling  under a barrier, we follow reference \cite{ScipostTunneling}, were they solve the one-dimensional scattering problem for the model
\begin{equation}
    \cal{H}=\begin{cases}
			(\frac{p_x^2}{2m}+V_0)\mathds{1}_{\sigma}+\Lambda p_x\sigma_y, & \text{if~ $0<x<a$}\\
            (\frac{p_x^2}{2m})\mathds{1}_{\sigma}, & \text{otherwise},
		 \end{cases}
		 \label{Hamiltonian2}
\end{equation}
where $\Lambda$ is the SO strength,$\sigma_y$ is the spin flip operator and $V_0$ the barrier height. $\mathds{1}_{\sigma}$ is the unit matrix in spin space and $\sigma_i$ are the Pauli spin matrices. $\cal{H}$ acts on the spinors $\psi=\left(\psi_+(x)~\psi_-(x)\right)$. This Hamiltonian contains only the spin active part of the helical model, so no orbital angular momentum is present\cite{VarelaZambrano,VarelaMujicaMedina2016,Oligopeptides2020}. Nevertheless, the magnitude of the SO coupling is derived from the overlaps pertaining to the chiral structure considered in those models. We emphasize that the above Hamiltonian should not be derived from the direct discretization of the Pauli Hamiltonian since $p_x$ is the crystal momentum of the electron, and the electric field in the Pauli Hamiltonian is that of the atomic nuclei. So, the transport SO coupling derives from a combination of atomic SO coupling and orbital overlaps in the molecular structure. Another important point is that chirality is built into the parameter $\Lambda$ i.e. chirality provides the inversion asymmetry for a momentum-dependent SO coupling as above around half filling, allowed by chiral symmetry.

An interesting analog that can be made with the SO term is that in analogy with Buttiker's $U(1)$ magnetic field, we can define a new momentum-dependent magnetic field $B_{\rm SO}$ given the mapping $\lambda p_x\sigma_y=-\gamma {\bf B}_{\rm SO}\cdot {\bm \sigma}$ that results in ${\bf B}_{\rm SO}=-(\Lambda/\gamma) p_x {\bf u}_y $. $B_{\rm SO}$ lies in the negative $y$ direction for the model Hamiltonian. So, one might expect that we have all the ingredients we need for a strong spin polarization in the direction of the $B_{SO}$ momentum-dependent magnetic field. 

Nevertheless, as even under the barrier, the wavevector is complex, unlike the magnetic field case, precession proceeds with no generation of a spin component along the ${\bf B}_{\rm SO}$ direction. Also, both spin components suffer the same decay within the barrier (although dependent on SO) independent of their spin orientation.  The equal treatment of both spin projections renders a null polarization. We can also see the "spin-orbit magnetic field" $B_{SO}$ does not perform the same role as the $U(1)$ magnetic field under the barrier since up and down spins components do not have different decay rates.

Another way to appreciate this result, using the same notation as in ref.\cite{ScipostTunneling}, is to see the incident up spin-oriented input state as an equal superposition of $|y_+\rangle$ and $|y_-\rangle$ states $(1 ~0)=1/{2}(1 ~i)+1/{2}(1~-i)$, eigenstates under the barrier. Computing then
\begin{equation}
P_y=\frac{|t_{+y}|^2-|t_{-y}|^2}{|t_{+y}|^2+|t_{-y}|^2}=\frac{2 \Im t^*_+ t_-}{|t_+|^2+|t_-|^2}=\langle s_y\rangle=0.
\label{PolarizationTransmission}
\end{equation}
The latter relations link the experimental and numerical formulas used in the literature for the polarization in terms of the expectation value of the spin\cite{ScipostTunneling}. This is the result expected for Bardarson's theorem for single channel two terminal models and, more generally, reciprocity\cite{Bart1,AharonyOraEntinComment}, since there is no source of time-reversal symmetry breaking as in the magnetic field case. The concocted ${\bf B}_{\rm SO}$ does not break TRS since it is momentum-dependent, as expected from the SO coupling.

The spin-orbit coupling treats spin projections equally, so it cannot account for polarized spin polarization as expected in the CISS effect alone.  Nevertheless, the conditions for reciprocity are {\it not met} if there is a coupling to a third probe beyond the two terminal setups. A thermalization of electron transport to the environment through the electron-phonon or electron-electron interactions is inevitable at room temperature, providing time-reversal symmetry breaking and completing the analogy to a $U(1)$ magnetic field\cite{ScipostTunneling,ButtikerProbe}. 

In the context of the model in ref.\cite{ScipostTunneling}, the environment can be modeled as a lumped probe that disrupts the delicate coherences that yield the reciprocity theorem\cite{Kiselev,Bardarson2008} in the linear regime. This turns our attention to a tunneling molecular system to a three-probe scenario following reference \cite{ScipostTunneling} using the Buttiker's voltage probe\cite{ButtikerProbe} generalized for spinors in reference~\cite{Ellner}. 

As a direct consequence, introducing this probe breaks time-reversal symmetry (TRS), giving rise to equivalent effects of a magnetic field, which induces a net spin polarization, in this case, parallel or antiparallel to $B_{\rm SO}$.
\begin{figure}
    \centering
    \includegraphics[width=8.5cm]{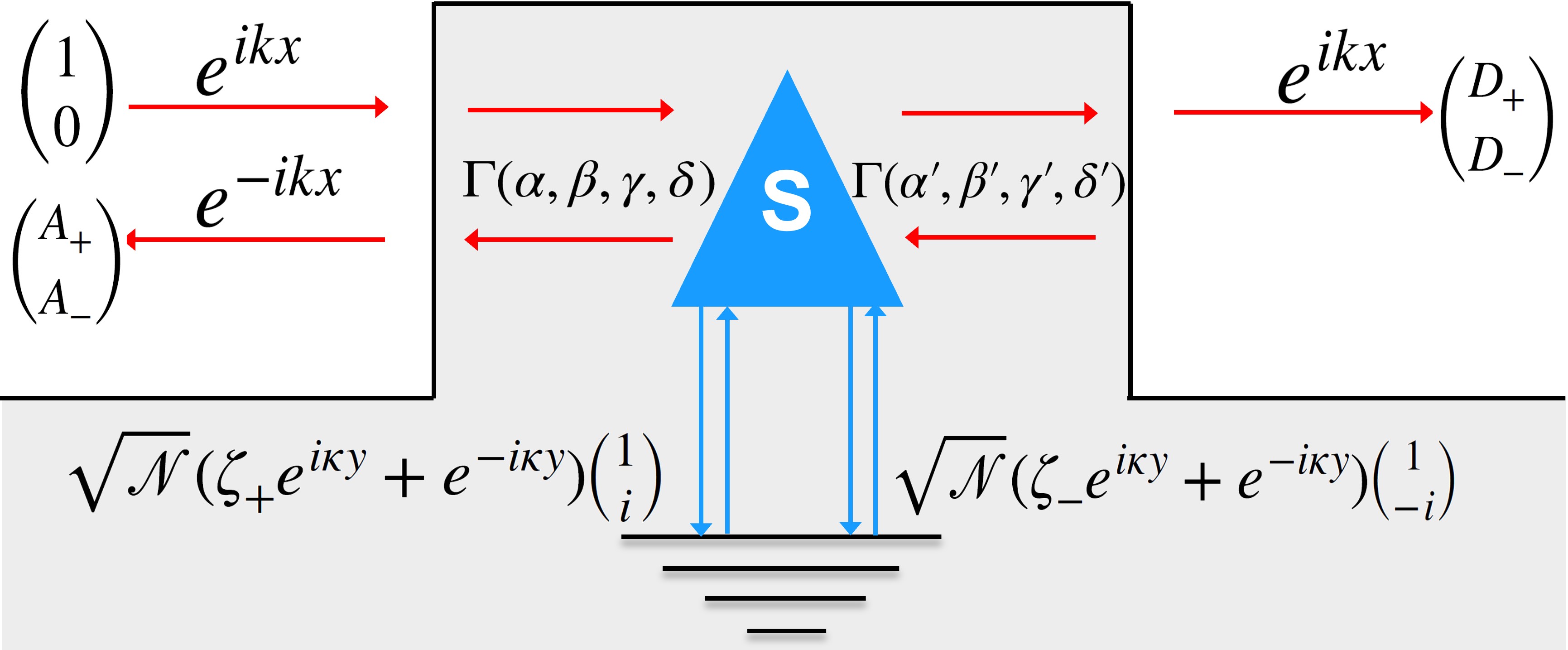}
    \caption{Buttiker's probe under the spin-orbit active barrier. The probe absorbs each eigenstate spin species under the barrier with the same scattering matrix, so no spurious spin selection is induced. Flux conditions are imposed on building the S matrix for a wideband Buttiker probe. Figure from reference\cite{ScipostTunneling}.}
    \label{ButtikerProbe}
\end{figure}

Figure \ref{ButtikerProbe} shows the four regions that must be matched for continuity and flux. Under the barrier, the matching occurs at position $(x_0,y_0)=(x_0,0)$ where $y$ describes the coordinate of the third probe. The Scattering ($S$) matrix can then emulate a generic dephasing process\cite{Ellner}.

\begin{figure}
    \centering
    \includegraphics[width=8.5cm]{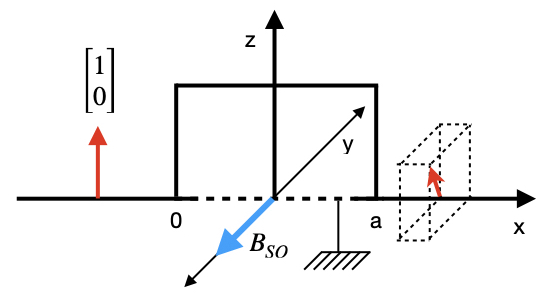}
    \caption{Spin tunneling setup chosen: We inject a spin-up electron in the $z$-axis into a spin-active barrier with a $B_{SO}$ in the $-{\hat y}$ direction. Asymmetric tunneling in the $y$ quantization axis produces a spin polarization or a net spin component in the $y$ direction. From ref.\cite{ScipostTunneling}.}
    \label{ButtikerPrecession}
\end{figure}

\begin{figure}
    \centering
    \includegraphics[width=8.5cm]{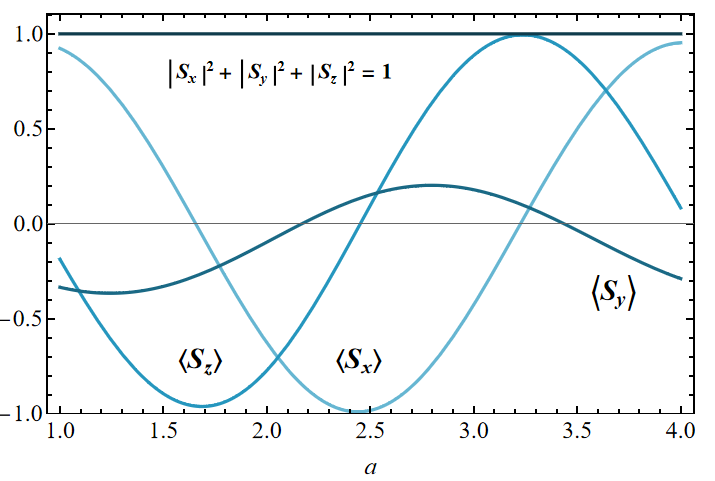}
    \caption{Spin precession when the decoherence probe couples to a particular point $x_0$ under the barrier. A disruption of spin precession in the $x-z$ plane is observed, generating a spin polarization analogous to an actual magnetic field\cite{ScipostTunneling} but with a relaxation direction depending on coupling strength and probe position. The norm of the spinor is preserved. The parameters depicted are $k=2~{\rm nm}^{-1}$, $k_{so}=1~{\rm nm}^{-1}$, $k_0=4~{\rm nm}^{-1}$, $a=1-4~{\rm nm}$, $s=1$, $x_0=0.8~{\rm nm}$, $\epsilon=0.02$, $\mathcal{N} = 0.01$. Plot computed with model in ref.\cite{ScipostTunneling}.}
    \label{ButtikerPrecessionDecoherence}
\end{figure}
\begin{figure}
    \centering
    \includegraphics[width=8.5cm]{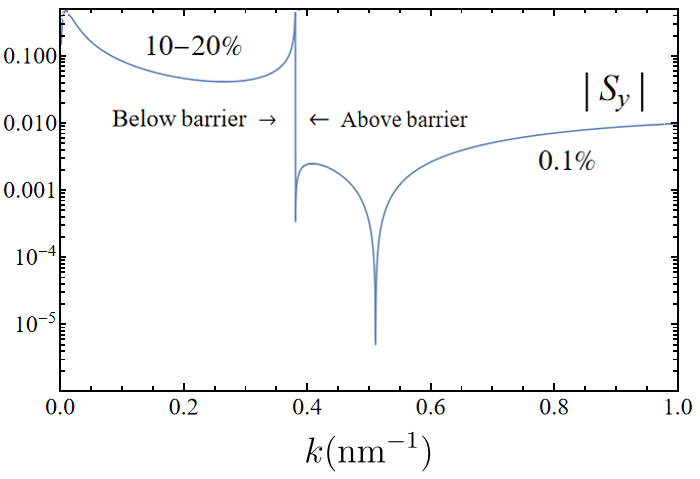}
    \caption{Spin Polarization as a function of the input wavevector $k$ with barrier height set at ($k_0=0.4~{\rm nm}^{-1}$) showing the difference in polarization of up to 3 orders for the polarization magnitude for the tunneling case versus polarization above the barrier.($k=0-1~{\rm nm}^{-1},\ k_{\rm so}=0.1~{\rm nm}^{-1},\ k_0=0.4~{\rm nm}^{-1},\ a=5~{\rm nm},\ s=1,\ x_0=1.5~{\rm nm},\ \epsilon=0.01,\ \mathcal{N} = 0.1$). Figure generated using the model of ref.\cite{ScipostTunneling}.}
    \label{PolarizationAboveBelowBarrier}
\end{figure}

\begin{figure}
    \centering
    \includegraphics[width=9cm]{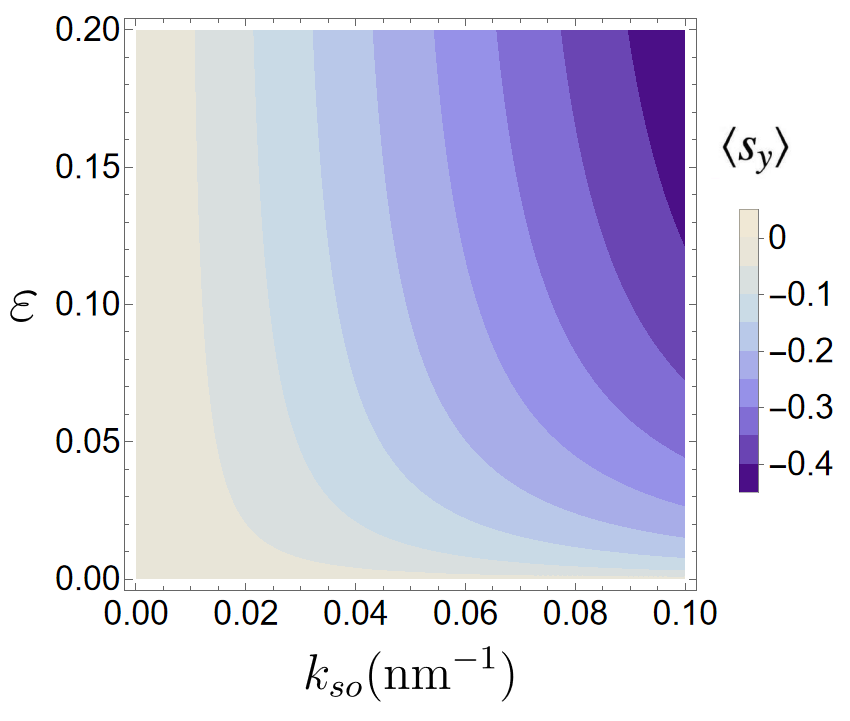}
    \includegraphics[width=9cm]{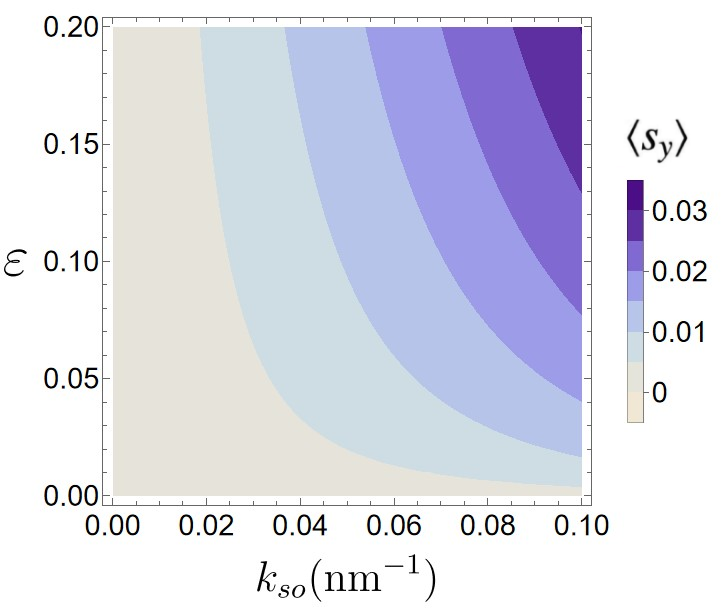}
    \caption{Top Panel: Spin polarization generated by decoherence analogous to that caused by a real external magnetic field. The contour plot shows the effect of a spin-orbit and decoherence coupling, consistent with estimates of ref.\cite{VarelaZambrano} for tunneling. The appearance of alignment of the spin to the $B_{\rm SO}$ is very sensitive to the coupling to the Buttiker probe. The parameter for under barrier: ($k=0.4~{\rm nm}^{-1},\ k_0=1~{\rm nm}^{-1},\  a=5~{\rm nm},\ s=1,\ x_0=1.5~{\rm nm},\  \mathcal{N} = 0.01$). Bottom Panel: For transmission above the barrier, the polarization is an order of magnitude weaker. The parameters here are ($k=0.4~{\rm nm}^{-1}$, $k_0=0~{\rm nm}^{-1}$, $a=5~{\rm nm}$, $s=1$, $x_0=1.5~{\rm nm}$,  $\mathcal{N} = 0.01$). Figure generated using model of ref.\cite{ScipostTunneling}.}
    \label{ContourEpsilonKsoAboveBelowBarrier}
\end{figure}
The spin scattering setup is depicted in Fig.\ref{ButtikerPrecession}, as in the coherent tunneling scenario, one injects spin up (in the $z$ quantization axis), which is decomposed into equal components in the basis functions of the SO term. 
Beyond the boundary conditions discussed at $x=0$ and $x=a$, we must include the matching equations at point $x=x_0$. One can solve for this system exactly\cite{ScipostTunneling}. The equal treatment of both $y$ and $-y$ projections according to Eq.[\ref{PolarizationTransmission}] is now broken, yielding a ``relaxation" toward the $y$ axis, as depicted in Fig.\ref{ButtikerPrecessionDecoherence}. This is analogous to what a magnetic field would achieve\cite{Buttiker,ScipostTunneling}.

Figure \ref{ButtikerPrecessionDecoherence} shows the appearance of the $y$ component of spin as a function of the barrier length in the presence of the decoherence probe. Note that polarization does not "relax" toward the particular ${\bf B}_{\rm SO}$ direction ($-y$) as in a magnetic field, as can be seen in the figure. The specific orientation on the ${\bf B}_{\rm SO}$ direction depends on the details of interference effects introduced by the probe\cite{ScipostTunneling}.

From the symmetry point of view, we have argued that breaking time-reversal symmetry is sufficient to bring about spin polarization, so, in the absence of tunneling or above the barrier height, we should still have spin polarization. 
Figure \ref{PolarizationAboveBelowBarrier} shows the magnitude of $s_y$ as one increases the input energy from below to above the barrier. While below the barrier, the polarization reaches between 10-20\%, above the barrier, the polarization is much lower by two orders of magnitude. The peaks in the figure correspond to the effects of precession and the energy passing the barrier height.

\begin{figure}
    \centering
    \includegraphics[width=9cm]{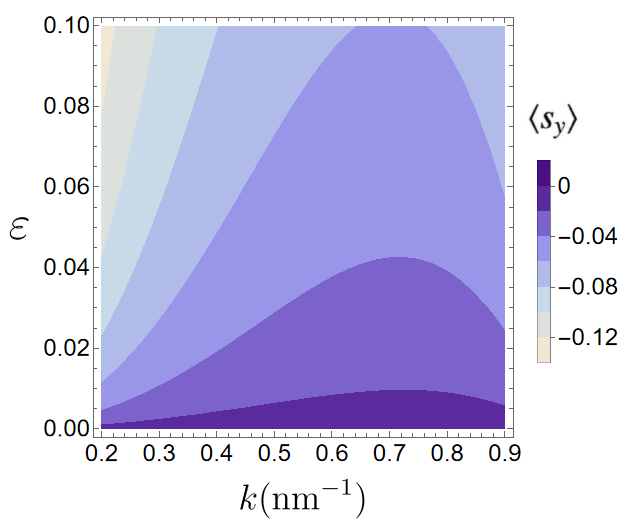}
    \includegraphics[width=9cm]{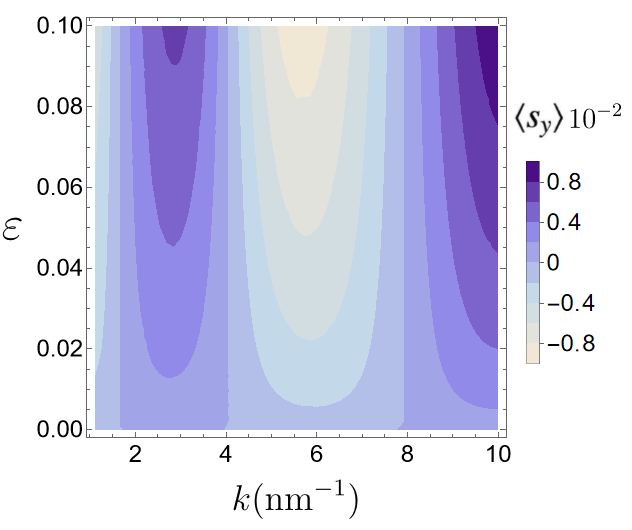}
    \caption{Top Panel: Spin polarization generated by decoherence under the barrier. Polarization is stronger and only antialigned to the field due to the tunneling effect. Bottom Panel: Spin polarization above the barrier. Polarization is weaker and oscillates between alignment and anti-alignment to the $B_{\rm SO}$ field. Here ($k_{so}=0.04~{\rm nm}^{-1},\ k_0=1~{\rm nm}^{-1},\ a=4~{\rm nm},\ s=1,\ x_o=0.8~{\rm nm},\ \mathcal{N} = 0.1$). Figure generated using the model of ref.\cite{ScipostTunneling}.}
    \label{ContourEnergyDependence}
\end{figure}

Figure \ref{ContourEpsilonKsoAboveBelowBarrier} compares the polarization power of scattering with barrier and without barrier as a function of the coupling to the Buttiker probe and the strength of the SO coupling. Finally, Figure\ref{ContourEnergyDependence} shows the polarization strength for scattering above and below the barrier as a function of the coupling strength to the Buttiker probe and the input energy. As previously discussed, one can see that while both situations polarize spin, energies below the barrier polarize orders of magnitude higher. Both cases can polarize spin as an interference effect that can change direction depending on the details of the input energy and coupling to the third probe.

\section{Discussion and Summary}

We have reviewed here the critical ingredients of the CISS effect assuming that it is an intrinsic effect of chiral molecules at non-zero temperatures. This is by no means the general perception in the community, where some argue that there is no effect in the absence of terminals/transport setup\cite{EversReview}. For this reason, we have focused on universal interactions that will be present in all systems claiming to exhibit CISS. We have also attempted to sow together the contributions for gas phase electron scattering off chiral molecules, where only forward scattering is polarized to photoelectron scattering off chiral self-assembled monolayers and two terminal transport measurements. We find that chirality and SO coupling are common once one discusses the nature of the polarization measured and the magnitude reported. The discussion also unifies point chirality, configuration chirality, and `helical' chirality as common resources in CISS. Configuration chirality, explained theoretically by Blum and Thompson\cite{BlumThompson1997} clarifies some claims that chirality is not a necessary ingredient.

For point chirality systems, chirality seems to enable the SO coupling, which is intrinsically inversion asymmetric, i.e., requires sources of inversion asymmetry to operate. This is also true for helical macromolecules where the atomic SO operates between nearest neighbors due to resulting chiral molecule orbital overlaps. Inversion asymmetry can also be provided by especially strong internal electric fields that, combined with atomic SO, yield Rashba spin active couplings\cite{Oligopeptides2020}.

As to the spin-orbit interaction itself, the meV scale of the intrinsic atomic coupling seems to be the only reasonable spin active resource for the experimental systems reported. This is also the case for spin activity in semiconductors, a very well-developed field\cite{Winkler}. For particular molecular structures, there remains to explore the so-called SIA (structural inversion asymmetry) and BIA (bulk inversion asymmetry) sources of SO in the language of semiconductor spintronics\cite{Winkler}. The current SO considered for CISS is of the SIA type. In the transport setup, there have been claims of the importance of the exchange interaction coupling the contact to the chiral molecule; see, for example, ref.\cite{ExchangeNaaman} as an integral part of CISS.

After the eye-opening work of ref.\cite{Bart1}, it became clear that a source of time-reversal symmetry breaking or effects in the non-linear transport regime was necessary. This is to observe two terminal spin-polarization or spin-induced magnetoresistance. Otherwise, linear regime reciprocity was to be observed (see also \cite{Huisman}). As CISS seems to exhibit spin-polarization in the linear regime\cite{ACSNanoReview} a source of time-reversal symmetry breaking seems in order. Decoherence seems to be that resource that breaks reciprocity in our view and is very clearly manifest in the earliest models\cite{GuoSunPRL,Matiyahu2016} albeit only with norm leakage models. We have reviewed how decoherence effects play a role in breaking reciprocity either from direct electron dephasing effects due to coupling to a third effective probe or inelastic scattering mediated by interactions such as electron-phonon or electron-electron. The corresponding thermal reservoir of vibrations of other thermalized electrons must accompany such interactions. 

Finally, after considering all the previous resources, matching the experimental numbers for spin polarization has baffled theory for many years. We show that tunneling can explain the numbers by combining it with decoherence in an exactly solvable model. Tunneling exponentiates asymmetries between spinor components produced by decoherence, producing large polarizations. The asymmetries depend on decoherence occurring in restricted length scales compared to the precession length or coherent spin-flipping events being more frequent than spin-dephasing events. The spin polarizing effect is due to interference, breaking the polarization-canceling effect of reciprocity. Thus, it does not choose a particular orientation with respect to the spin-orbit magnetic field ${\bf B}_{\rm SO}$.

To establish CISS as a standalone effect of chiral molecules, one must identify its manifestations in contexts without terminals or only weak interactions of chiral molecules and substrates. Although we have not specified how the ingredients we compiled in this review and how these manifestations play out, emblematic experiments suggest it. The first experiment was described in ref.\cite{NaamanEnantiomers}, where a racemic mixture of chiral species was demonstrated to be differentially attracted to a magnetized surface. This differentiated attraction led to one chirality staying attached to the surface and the other being washed away, providing a much-needed new method to separate chiralities. The preliminary explanation for the effect is a CISS-related spin-polarized ordering of the ends of chiral molecules that determined the interaction with the magnetized surface. The fact that the specific polarization of the molecules that can explain this separation method is related to the in sources we have pointed out has not yet been addressed.

Another recent emblematic experiment of ref.\cite{Eckvahl} claimed the ``pure" CISS effect in isolated covalent donor-chiral bridge-acceptor molecules. This is an electron transfer between two sites in a molecule, as we described before. The claim is that CISS strongly influences the spin dynamics of isolated covalent donor-chiral bridges. The described results involved a degree of quantum behavior. They added decoherence and established the relevance of chirality and that the substrates or electrodes with their possibly large spin-orbit couplings are unnecessary for CISS to occur.

Finally, another recent proposal that explored CISS in isolated chiral molecules is that of ref.\cite{SolmarChiroOptical}, where they studied the optical rotation power of chiral molecules coupled to the SO and the asymmetry of the Kramers pairs induced by decoherence. They found the analog of reciprocity without decoherence through the absence of spin-related circular dichroism. In contrast, decoherence brings specific signatures thereof, such as chiral asymmetries as a function of the incident wavelengths.

We hope this review will help establish a unified theory for CISS and further develop it into a fundamental resource for molecular spintronics applications.

\section*{Acknowledgments}
M.M, acknowledges the hospitality of The Chair in Material Science and Nanotechnology, TU Dresden, where part of this work was done. E.M. Acknowledges support from Poligrant 17617 USFQ and an invited professorship at Lorraine University, where this work began. B.B. Acknowledges support from LA-CoNGA physics (Latin American Alliance for Capacity buildiNG in Advanced Physics).


\clearpage

\end{document}